\documentstyle[aps,pre,multicol]{revtex}
\begin{document}
\def\nn {\nonumber}
\def\OP{\tensor P}
\def\B.#1{{\bbox{#1}}}
\def\BE {\begin{equation}}
\def\EE {\end{equation}}
\def\BEA {\begin{eqnarray}}
\def\EEA {\end{eqnarray}}
\def\Fbox#1{\vskip0ex\hbox to 8.5cm{\hfil\fboxsep0.3cm\fbox
{\parbox{8.0cm}{#1}}\hfil}\vskip0ex}
\def\al {\B.{A}}
\def\w{\B.{\cal W}}
\def\v{\B.{\cal V}}
\def\l{\hat {\cal L}}
\def\r{\hat {\cal R}}
\def\p{\hat P}
\def\q{\hat Q}
\def\k{\hat J}
\def\t{\hat I}
\def\cP{\B.{\cal P}}
\def\a {\alpha}
\def\b{\beta}
\def\g{\gamma}
\title{Continued Fraction Representation of Temporal Multi Scaling in
  Turbulence}
\author {David Daems$^{*,**}$, Siegfried Grossmann$^{\dag}$, Victor S.
  L'vov$^{*}$ and Itamar Procaccia$^{*}$}
\address{$^{*}$Department of~~Chemical Physics,
  The Weizmann Institute of Science, Rehovot 76100, Israel\\
  $^{**}$Center for Nonlinear Phenomena and Complex Systems, Universit\'e
  Libre de Bruxelles, 1050 Brussels, Belgium\\
  $^{\dag}$Fachbereich Physik,   Philipps Universitaet Marburg,
  Renthof 5, Marburg D-35032, Germany}
\maketitle
\begin{abstract}
  It was shown recently that the anomalous scaling of simultaneous
  correlation functions in turbulence is intimately related to the
  breaking of temporal scale invariance, which is equivalent to the
  appearance of infinitely many times scales in the time dependence of
  time-correlation functions.  In this paper we derive a
  continued fraction representation of turbulent time correlation
  functions which is exact and in which the multiplicity of time
  scales is explicit. We demonstrate that this form yields precisely
  the same scaling laws for time derivatives and time integrals as the
  ``multi-fractal" representation that was used before. Truncating the
  continued fraction representation yields the ``best" estimates of
  time correlation functions if the given information is limited to the
scaling exponents of the simultaneous correlation functions up to a
certain, finite order.
It is worth noting that the derivation of a continued fraction representation
obtained here for an operator which is not Hermitian or anti-Hermitian 
may be of independent interest.
\end{abstract}
\begin{multicols}{2}
\section{Introduction}
It is commonly argued \cite{97SA,Fri} that fully developed
hydrodynamic turbulence exhibits simultaneous statistical objects
whose scaling properties are anomalous.  For example the so called
structure functions satisfy scaling laws of the form
\begin{equation}
S_n(R)=\left<|\B.u(\B.r+\B.R,t) -\B.u(\B.r,t)|^n\right> \sim R^{\zeta_n} \
, \label{Sn}
\end{equation}
where $\B.u(\B.r,t)$ is the Eulerian velocity field, and $\zeta_n$ are
scaling exponents which are nonlinear functions of $n$. The nonlinear
dependence is referred to as ``anomalous scaling" or ``multi-scaling",
and the issue of evaluating these exponents from either
phenomenological models or from first principles has attracted
significant amount of effort in the last decade.

It has only recently been discovered \cite{97LPP} that also the time
dependence of the $n$-th order correlation functions is multi-scaling,
and that ``dynamical scaling" is broken.  This phenomenon seems to
distinguish turbulence from other problems in which scaling is
anomalous, like critical phenomena. In the latter case dynamical
scaling is invoked by stating that a space time correlation function
$F(R,t)$ is a homogeneous function of its arguments in the sense that
$F(\lambda R,\lambda^z t)=\lambda^\zeta F(R,t)$, where $\zeta$ and $z$
are the ``static" and ``dynamic" scaling exponents respectively.  In
turbulence such relations do not exist even when the same-time
correlation functions are homogeneous functions of the spatial
coordinates. The importance of this fact in determining the structure
of the theory has been stressed in \cite{98LP}, and see also
\cite{98BLPP}.

In this paper we address temporal multi-scaling on the basis of the
continued fraction representation of turbulent correlation function
\cite{82GT,97GW}. This approach will lead us to a different point of view of
temporal multi-scaling, in agreement with the conclusions as
Ref.\cite{97LPP}. The advantage of the present formulation is
three-fold: first, it is derived on the basis of an exact formulation
of the time correlation functions and their dynamics. The phenomenon
of temporal multi-scaling is related in this approach to the scaling
properties of higher order temporal derivatives of correlation
functions, computed at zero time. Second, this approach furnishes
information not only about leading scaling exponents, but also about
sub-leading ones. Third, a finite truncation of the continued fraction
representation is in a sense the ``best" possible representation when
the information about the scaling of one-time correlation functions is
limited to the low order scaling exponents.
We will show that the scaling laws exhibited by the continued fraction
representation are identical to those predicted by the ``multi-fractal"
representation of time-correlation function\cite{98LP}, adding weight
and justification to the latter. Since the multi-fractal representation
was used recently to estimate the scaling exponents from first
principles \cite{98BLPP}, we ascribe some weight to being able to
justify it further.

To keep the formalism minimal and the result clearest, we treat in
this paper only the second order space-time correlation function of
turbulent fields. The formalism can be used to generate
representations of any higher order correlation function, but we do
not elaborate on this in the present text. In Section II we review
briefly the Zwanzig-Mori formalism\cite{61Zwa,65Mor} which was applied
to time correlation functions in turbulence \cite{82GT,97GW}, and display
the continued fraction representation of the second order time
correlation function of the velocity field.  We show that the
coefficients in this representation can be written in terms of time
derivatives at zero time of the same second order correlation
function. In Section~III we set up a theory for the evaluation of the
scaling exponents of these coefficients. In Section~IV we derive the
scaling laws implied by the continued fraction representation
for the time derivatives of the correlation
function evaluated at time zero.
In Section~V
 we relate our
results to the multi-fractal representation of the correlations
functions, and explain the scaling-equivalence of the two
representations. In Section~VI we offer a summary and a short
discussion.
\section{continued fraction representation}
In thinking about dynamics one cannot deal with time-correlation
functions of the Eulerian field since these are dominated by the
kinematic sweeping time scale. We need to consider Lagrangian or
Belinicher-L'vov velocity differences. We prefer the latter since they
obey Navier-Stokes like equations of motion which are local in time.
In terms of the Eulerian velocity ${\B.u}({\B.r},t)$ Belinicher and
L'vov defined\cite{87BL} the field ${\B.v}({\B.r}_0,t_0\vert
{\B.r},t)$ as
 \begin{equation}
 {\B.v}({\B.r}_0,t_0\vert {\B.r},t)\equiv {\B.u}\lbrack{\B.r}
 +\mbox{\boldmath$\rho$}
 ({\B.r}_0,t),t\rbrack
 \label{a2}
 \end{equation}
 where
 \begin{equation}
 \mbox{\boldmath$\rho$}  ({\B.r}_0,t)
  =\int_{t_0}^{t} ds {\B.u }[{\B.r}_0 +\mbox{\boldmath$\rho$}({\B.
  r}_0,s) ,s] \ .
  \label{a3}
 \end{equation}
The observation of Belinicher and L'vov \cite{87BL} was that the
variables
$\B.{\cal
   W}(\B.r_0,t_0|\B.r,\B.r,t)$ defined as
 \begin{equation}
\B.{\cal W}(\B.r_0,t_0|\B.r,\B.r',t)\equiv
\B.v( {\B.r}_0,t_0|{\B.r},t)-
\B.v( {\B.r}_0,t_0|\B.r',t)  \ , \label{newBL}
\end{equation}
exactly satisfy a Navier-Stokes like equation in the incompressible
limit:
\end{multicols}
\vskip -0.4cm
\leftline{----------------------------------------------------------------------
 --}
\begin{equation}
\Big[{\partial\over \partial t}+\tensor{\B.\cP} \B.{\cal
 W}({\B.r}_0,t_0|{\B.r},\B.r_0,t)
\cdot{\B.\nabla_r}+
\tensor{\B.\cP}' \B.{\cal W}({\B.r}_0,t_0|{\B.r'},\B.r_0,t)
\cdot{\B.\nabla_r'}-\nu (\nabla_r^2+\nabla_r'^2)\Big]
\B.{\cal W}(\B.r_0,t_0|\B.r,\B.r',t)=0\ .
\label{newNS}
\end{equation}\vskip .1cm
\rightline{---------------------------------------------------------------------
 ---}

 \begin{multicols}{2}
   We remind the reader that the application of the transversal
   projector ${\tensor{\B.\cP} }$ to any given vector field $ {\B.a}(
   {\B.r})$ is non local, and has the form:
 \begin{equation}
 \lbrack\tensor{\B.\cP}{\B.a}(
 {\B.r})\rbrack^\alpha =\int d  \tilde\B.r {\cal P}^{\alpha\beta}(
 {\B.r}-\tilde\B.r)a^\beta(\tilde\B.r).
 \label{b2}
 \end{equation}
 The explicit form of the kernel ${\cal P}^{\alpha\beta}(\B.r)$ can be
 found, for example, in \cite{MY}.  In (\ref{newNS}) $\tensor{\B.\cP}$
 and $\tensor{\B.\cP}'$ are projection operators which act on fields
 that depend on the corresponding coordinates $\B.r$ and $\B.r'$.  For
 our purpose below we introduce the Liouville operator $\hat{\cal
   L}(\B.r_0,t_0|\B.r,\B.r')$ which is defined via the total time
 derivative of $\B.{\cal W}(\B.r_0,t_0|\B.r,\B.r',t)$ at time $t=t_0$:
\begin{equation}
{d \B.{\cal W}(\B.r_0,t_0|\B.r,\B.r',t)\over dt}\Bigg|_{t=t_0}
\!\!\!\!\! \!\!\equiv
\hat{\cal L}(\B.r_0,t_0|\B.r,\B.r') \B.{\cal W}(\B.r_0,t_0|\B.r,\B.r',t_0)\,.
\end{equation}
The identity of the Liouville operator follows by definition,
\begin{eqnarray}
&&{d{\B.v}({\B.r}_0,t_0\vert \B.r-\mbox{\boldmath$\rho$}
 ({\B.r}_0,t),t\rbrack,t)\over dt}\equiv {d{\B.u}(\B.r,t)\over dt}
\nonumber\\&=& {\partial \B.u(\B.r,t)\over \partial t}+[\B.u(\B.r,t)\cdot
\B.\nabla]\B.u(\B.r,t)
\end{eqnarray}
Translating all coordinates by $\mbox{\boldmath$\rho$}({\B.r}_0,t)$ we
find after elementary algebra,
\begin{eqnarray}
\hat{\cal L}(\B.r_0,t_0|\B.r,\B.r')={\partial\over \partial
t}\Bigg|_{t=t_0}&+&\B.v( {\B.r}_0,t_0|{\B.r},t_0)
\cdot\B.\nabla_r\nonumber\\&+&\B.v(
{\B.r}_0,t_0|{\B.r'},t_0)\cdot\B.\nabla_{r'}\ .
\end{eqnarray}
Consider now the time dependence of ``fully fused" 2nd order
correlation function
\begin{equation}
{\cal F}_2^{\a\b}(\B.r_0|\B.r,\B.r',\tau)
= \left<{\cal W}^\a(\B.r_0|\B.r,\B.r',t_0){\cal
 W}^\b(\B.r_0|\B.r,\B.r',t_0+\tau)  \right> .
\label{defF2t}
\end{equation}
By ``fully fused" we mean here that the coordinates of the two
velocity differences are the same, and they differ only in their time
argument. The same-time counterpart of this correlation function, i.e.
$\B.{\cal F}_2(\B.r_0|\B.r,\B.r',\tau=0)$ is independent of $\B.r_0$
and in an isotropic and homogeneous ensemble it is a function of
$|\B.r-\B.r'|$ only.
Accordingly it differs from the standard structure function
$S_2(|\B.r-\B.r'|)$ only in having the full 2nd rank tensorial
character.  For this analysis we choose all the three vector distances
to have the modulus in the inertial range, of the order of $R$. In the
notation we will keep only this $R$ but remember that the angular
dependence is not shown explicitly.

To proceed, we consider the tensorial correlation function
(\ref{defF2t}) as an inner product in the space of vectors
$\B.{\cal W}$, denoted as
\BE \B.{\cal F}_2(R,\tau)=\left(\w ,e^{\l \tau}
\w\right) \ . \label{expL}
\EE
In particular we are interested in the Laplace transform of
Eq.~(\ref{expL})
\BE
\label{Fz}
\tilde \B.{\cal F}_2(R,z)=(\w, \frac{1}{z-\l} \w)\ .
\EE
It has been shown by Grossmann and Thomae \cite{82GT} that the
Zwanzig-Mori projection operator formalism \cite{61Zwa,65Mor} applies
to turbulent systems described by Navier-Stokes like equations.  In \cite{97GW}
it has been demonstrated that the contribution of the memory kernel and
higher order continued fraction is considerable. The
central idea is to decompose the resolvent
\BE
\r(z)=\frac{1}{z-\l}
\EE
by means of a projection operator $\p$.  With $\q=\hat 1-\p$ as
projector orthogonal to $\p$ one has the resolvent identity
\BE
\label{resolv}
\p\r(z)\p=\p\frac{1}{z-\p\l\p-\p\l\q\frac{\displaystyle 1}{ \displaystyle
 z-\q\l\q}\q\l\p}\ .
\EE
Since we note that the time correlation function of interest is
the $\w-\w$ matrix element of the resolvent, it is useful to choose
$\p$ to be the projector on $\w$,
\BE
\p . \equiv (\w,.)(\w,\w)^{-1}\w \ .
\EE
Its basic properties are $\p\p=\p$ (idempotent) and $\p^{\dag}=\p$
(self adjoint), characterizing an orthogonal projection.
>From Eq.(\ref{resolv}) we have an expression for $\p$:
\BE
\p=\p\r(z)\p\left[z-\p\l\p-\p\l\q\frac{1}{z-\q\l\q}\q\l\p\right]
\EE
Compute now the $\w-\w$ matrix element, use $\p\w=\w$, as well as the
definition of
$\p$ to compute
\BEA
&&(\w,\r(z)\w)(\w,\w)^{-1}[z(\w,\w)\\
&&-(\w,\l\w)-(\w,\l\q \frac{1}{z-\q\l\q}\q\l\w)]=(\w,\w)\nonumber \ .
\EEA
We therefore conclude with
\BE
\tilde\B.{\cal F}_2(R,z)=\frac{\B.k_0(R)}{z-\B.\gamma_0(R)-\tilde\B.K_0(R,z)} \
, \label{step0}
\EE
where
\BEA
&&\B.k_0(R)=(\w,\w)\,,\nonumber\\ \label{coef0}
&&\B.\gamma_0(R)=(\w,\l\w)(\w,\w)^{-1}\,, \\
&&\tilde\B.K_0(R,z)=\Big(\q^{\dag}\l^{\dag} \w,\frac{1}{z-\q\l\q}\q\l\w\Big)
(\w,\w)^{-1}\ . \nonumber
\EEA
Here, of course, $\q^{\dag}=\q$, but in the next steps of generating
the continued fraction hermiticity will not hold.
Realizing that the kernel $\tilde\B.K_0(R,z)$ has the
same resolvent structure as $\tilde\B.{\cal F}_2(R,z)$ except that it
now features $\q\l\q$ instead of $\l$ and the states are
$\q^{\dag}\l^{\dag}\w$, $\q\l\w$ instead of $\w$, $\w$, with $\l^{\dag}$
the adjoint of $\l$ - one can continue the
fraction by the same procedure.
This is more transparent if we denote
\BE
\w_1=\q\l\w\,,\quad \tilde\w_1=\q^{\dag}\l^{\dag}\w\,,\quad \l_1=\q\l\q
\EE
so that $\tilde\B.K_0(R,z)$ takes on the form
\BE
\label{K0z}
\tilde\B.K_0(R,z)=\Big(\tilde\w_1,\frac{1}{z-\l_1}\w_1\Big)(\w,\w)^{-1}
\ .
\EE
Now we define a new projection operator
\BE
\p_1 \,\cdot \equiv (\tilde\w_1,\cdot)(\tilde\w_1,\w_1)^{-1}\w_1
\ .
\EE
As a result of $\tilde\w_1$ being different from $\w_1$
when $\l$ is not Hermitian or anti-Hermitian, this operator $\p_1$
is not Hermitian and performs accordingly non-orthogonal projections.
But $\p_1$ still is idempotent, $\p_1\p_1=\p_1$, which is the essential
property for deriving the analogous resolvent identity with $\l_1$ in
(\ref{K0z}) as for the original resolvent (\ref{Fz}) with $\l$.
Defining $\q_1 \equiv 1-\p_1$ we can repeat the argument leading to
(\ref{step0}) and (\ref{coef0}), and find
\BE
\tilde\B.K_0(R,z)=\frac{\B.k_1(R)}{z-\B.\gamma_1(R)-\tilde\B.K_1(R,z)} \ ,
\label{step1}
\EE
where
\BEA
&&\B.k_1(R)=(\tilde\w_1,\w_1)(\w,\w)^{-1}\,,\\
&&\B.\gamma_1(R)=(\tilde\w_1,\l_1\w_1)(\tilde\w_1,\w_1)^{-1}\,,
 \label{coef1}\nonumber\\
&&\tilde\B.K_1(R,z)=\Big(\q_1^{\dag}\l_1^{\dag}\tilde\w_1,
\frac{1}{z-\q_1\l_1\q_1}\q_1\l_1\w_1
 \Big )\nonumber \\
&&\qquad\qquad \quad \times (\tilde\w_1,\w_1)^{-1} \ .
 \nonumber
\EEA
Hence starting from
\BE
\label{Knz}
\tilde\B.K_n(R,z)=\Big(\tilde\w_{n+1},\frac{1}{z-\l_{n+1}}\w_{n+1}
\Big)(\tilde\w_{n},\w_{n})^{-1}
\ ,
\EE
with $n\geq 0$ one arrives at
\BE
\tilde\B.K_n(R,z)=\frac{\B.k_{n+1}(R)}{z-\B.\gamma_{n+1}(R)-\tilde\B.K_{n+1}(R,z
 )} \ ,
\label{stepn+1}
\EE
where
\BEA
\label{defk}
&&\B.k_{n+1}(R)=(\tilde\w_{n+1},\w_{n+1})(\tilde\w_n,\w_n)^{-1} \ ,\nonumber\\
&&\B.\gamma_{n+1}(R)=(\tilde\w_{n+1},\l_{n+1}\w_{n+1})(\tilde\w_{n+1},\w_{n+1})^
 {-1}\ .
\EEA
Here we used the notation
\BEA
\label{defWL}
&&\w_{n+1}=\q_n\l_n\w_n\,,\
\tilde\w_{n+1}=\q_n^{\dag}\l_n^{\dag}\tilde\w_n\,,  \nonumber\\
&&\l_{n+1}=\q_n\l_n\q_n\,, \ \w_{0}=\tilde\w_{0}=\w\,, \
\l_{0}=\l\,,
\EEA
and defined new projection operators as
\BEA
\p_{n}\,\cdot&=&
(\tilde\w_{n},\cdot)(\tilde\w_{n},\w_{n})^{-1}\w_{n}\nonumber\\
\q_{n}&=&\hat 1- \p_{n} \ .
\label{def}
\EEA
From (\ref{step0}) and (\ref{stepn+1}) it thus follows that the Laplace
transform $\tilde\B.{\cal F}_2(R,z)$ of the correlation
function (\ref{defF2t}) can be written in continued fraction representation:
\begin{equation}
\tilde\B. {\cal F}_2(R,z)=\frac{\B.k_0(R)}{\displaystyle z-\B.\gamma_0(R)
-\frac{\B.k_1(R)}{\displaystyle z-\B.\gamma_1(R)
-\frac{\B.k_2(R)}{\displaystyle z-\B.\gamma_2(R)-\ddots}}} \ . \label{contfrac}
\end{equation}
The novelty when the
operator $\l$ is not Hermitian or anti-Hermitian as is the case here
is that the new projection operators introduced to continue the
fraction perform non-orthogonal
projections.  Only the lowest order iterates of $\p_n$ and
$\q_n$, i.e. the operators $\p$ and $\q$, perform orthogonal
projections.
In the following section we analyze the scaling properties of this
 representation.

\section{scaling properties of the continued fraction representation}
In this section we determine the {\em leading} scaling exponents of
the coefficients which appear in the continued fraction representation
(\ref{contfrac}).  This leading scaling behavior is given in terms of
correlation functions of time derivatives of the velocity differences
$\B.{\cal W}$ computed at zero time:
\BEA
\label{kgs}
&&\B.k_{n}(R) \approx
\langle\l^{n}\w \l^{n}\w \rangle/
\langle\l^{n-1}\w \l^{n-1}\w \rangle \,, \\
&&\B.\gamma_{n}(R) \approx
\langle\l^{n+1}\w \l^{n}\w \rangle/
\langle\l^{n}\w \l^{n}\w  \rangle  \ . \nonumber
\EEA
Here the symbol $\approx$ means ``leading scaling order". Eq.(\ref{kgs})
yields the following explicit scaling
\BEA
\B.k_0&\propto& R^{\zeta_2} \ , \label{scalekg}\\
\B.k_n(R)&\propto& R^{\zeta_{2n+2}-\zeta_{2n}-2} \ , \quad n\ge1
\nonumber\\
\gamma_n(R)&\propto& R^{\zeta_{2n+3}-\zeta_{2n+2}-1} \ . \nonumber
\EEA
The rest of this section is a demonstration of this result. The reader who
prefers to see the connection to the multi-fractal representation can go
 directly
to the next section.

Let us define
\BE
\label{defjn}
\k_n\equiv\q_0\ldots\q_{n-1}\q_n\q_{n-1}\ldots\q_0=\q_n \ .
\EE
Now
\BE
\p_i\p_j=\delta_{ij}\p_i \ , \ \ \ \ \ i,j \geq 0
\EE
implies that
\BEA
\label{prok}
\k_n&=&\q_0\q_1\ldots\q_n=\q_n\q_{n-1}\ldots\q_0=\q_n\nn\\&=&1-\p_0-\ldots-\p_n \ .
\EEA
Hence Eq. (\ref{defWL}) leads to
\BEA
\w_{n+1}&=&\k_n\l \k_{n-1}\l \k_{n-2}\ldots\l \k_0\l\w\,, \nonumber\\
\tilde\w_{n+1}&=&\k_n^{\dag}\l^{\dag} \k_{n-1}^{\dag}\l^{\dag}
\k_{n-2}^{\dag}\ldots\l^{\dag}
 \k_0^{\dag}\l^{\dag}\tilde\w\,, \nonumber\\
\l_{n+1}\w_{n+1}&=&\k_n\l \k_n\l \k_{n-1}\ldots\l\k_0\l\w \ .
\EEA
It follows that
\BEA
&&(\tilde\w_{n+1},\w_{n+1})\nonumber\\
&=&(\w,\l\k_0\l\k_{1}\ldots\l\k_{n-1}\l\k_{n}\l\k_{n-1}\ldots\l\k_0\l\w)\,,
\nonumber\\
&&(\tilde\w_{n+1},\l_{n+1}\w_{n+1})\nonumber\\
&=&(\w,\l\k_0\l\k_{1}\ldots\l\k_{n}\l\k_{n}\l\k_{n-1}\ldots\l\k_0\l\w)
\ .
\label{lak}
\EEA
In order to arrive at Eq. ~(\ref{kgs}) we have to show that
\BEA
&&(\tilde\w_{n+1},\w_{n+1}) \approx (\w,\l^{2n+2} \w)\nonumber \\
\label{lak2}
&&(\tilde\w_{n+1},\l_{n+1}\w_{n+1})  \approx (\w,\l^{2n+3}\w)\ .
\EEA
It suffices to use the following assertion  which we prove below
\BE
\label{sup}
(\w,\l^{k}\k_{q}\t_q\w) \approx (\w,\l^{k}\t_q\w)\,,  \qquad  k>q\geq 0\,,
\EE
where $\t_q$ is any composition of $\l$ and $\k_i$ which has two
 properties.
First, it contains at
least $q+1$ times the operator $\l$ which we indicate by the subscript
 $q$ in $\t_q$ and second, on the right hand of each $\k_i$ there are
 at least $i+1$ operators $\l$ in the composition.
Since the compositions of $\l$ and $\k_i$ appearing  in (\ref{lak})
satisfy these two properties, (\ref{lak2}) results straightforwardly.

We now proceed to the proof of~(\ref{sup}) by induction.
Assuming for $p<q<k$ that
\BE
\label{hyp}
(\w,\l^{k}\k_p\t_p\w) \approx (\w,\l^{k}\t_p\w) \ ,
\EE
we show that
\BE
\label{con}
(\w,\l^{k}\k_q\t_q\w) \approx (\w,\l^{k}\t_q\w)  \ .
\EE
Using Eq. (\ref{prok}) the LHS of Eq. (\ref{con}) becomes
\BEA
\label{eq1}
(\w,\l^{k}\k_{q}\t_q\w)=(\w,\l^{k}\k_{q-1}\t_q\w)-
(\w,\l^{k}\p_q\t_q\w) \ . \nonumber\\
\EEA
By Eq. (\ref{hyp}) the first term on the RHS scales  as
\BE
\label{eq2}
(\w,\l^k\k_{q-1}\t_q\w) \approx (\w,\l^{k}\t_q\w) \ .
\EE
As for the second term, without specifying $\t_q$ we notice
using Eq. (\ref{def}) for $\p_q$
that it  is the product
of the following three factors that feature only
    operators
$\k_p$ with $p<q$ and with more than $p$ times the operator  $\l$           on
 their
right hand so that  by (\ref{hyp}) one has
\BEA
\label{eq3}
(\w,\l^{k}\w_q) &=&(\w,\l^k\k_{q-1}\l\k_{q-2} \ldots \l\k_0\l\w) \nonumber
\\&\approx&
(\w,\l^{k+q}\w)\,, \nonumber\\
(\tilde\w_q,\w_q)&=&(\w,\l\k_{0}\l\k_{1} \ldots\l\k_{q-1}\ldots \l\k_0\l\w)
\nonumber \\&\approx&
(\w,\l^{2q}\w) \,, \nonumber\\
(\tilde\w_q,\t_q\w)&=&(\w,\l\k_{0}\l\k_{1} \ldots\l\k_{q-1}\t_q\w)\nonumber
\\& \approx & (\w,\l^{q}\t_q\w) \ .
\EEA

In order to compare the scaling behavior of the two terms on the RHS of
Eq.~(\ref{eq1}) in the limit of infinite Reynolds number  we recall
that within the inner product (or
equivalently the average operation) an operator $\l$ amounts to simply $\w/R$
\cite{97LPP}. This follows from the convergence in the UV and in the IR (in
the limit Re$\to \infty$) of the integral implied by
the terms  ${\tensor{\B.\cP} }
\w \cdot{\B.\nabla_r}$ in Eq.~ (\ref{newNS}), so that the leading
contribution comes
from distances of the order of $R$. In other words, $\l$ in a correlation
function, when it operates on $\w$, introduces a term of the order of
$\w\cdot \B.\nabla \w$,
which, due to the demonstrated locality in scale space, can be estimated
as adding to the correlation a factor of the order of $\w/R$.
Hence one has
\BE
\label{prescription}
(\w,\l^{k}\w)\approx \left< \frac{\w^{k+2}}{R^k} \right> \propto
 R^{\zeta_{k+2}-k}\ ,
\EE
where the last step follows from Eq. (\ref{Sn}).

The composition $\t_q$ containing say $j^*$ times the operator  $\l$
with $j^* > q$ and having at least $i$ times the operator $\l$ on the
right hand of each operator $\k_i$, one can write
\BE
\label{eq4}
\t_q\w=\sum_{j=0}^{j^*} c_j(R)\l^j\w \ ,
\EE
where $c_j(R)$ is a function of the separation distance $R$ resulting
from all the contributions to $\l^j\w$.
Note in particular  that $c_{j^*}(R)=1$.
Considering now the following quantity for arbitrary $s$ one has
\BEA
\label{eq5}
(\w,\l_s\t_q\w) \propto \sum_{j} c_j(R)R^{\zeta_{s+j+2}-j-s}\ .
\EEA
Because of the H\"older inequalities
\BE
\label{Holder}
\zeta_{m}   -\zeta_{m-r} \leq \zeta_{n}-\zeta_{n-r} \ \ \ \
\ m>n \ , \ r>0 \ ,\
\EE
to which any two contributions of Eq. (\ref{eq5}) can be reduced,
the leading contribution is that for
which the  index of $\zeta$ is maximum, i.e. $s+j^*+2$.
As a result
\BE
\label{eq6}
(\w,\l_s\t_q\w) \propto R^{\zeta_{s+j^*+2}-j^*-s}\ .
\EE
It follows from Eqs. (\ref{eq2}) and (\ref{eq6}) that the scaling
behavior of the first term on the RHS of Eq. (\ref{eq1}) is
\BEA
\label{eq7}
(\w,\l_k\t_q\w) \propto R^{\zeta_{k+j^*+2}-k-j^*}\ ,
\EEA
whereas using Eqs. (\ref{eq3}),(\ref{prescription}) and (\ref{eq6})
 that of the second one is
\BE
\label{eq8}
(\w,\l^k\p_q\t_q\w)\propto R^{\zeta_{k+q+2}-
\zeta_{2q+2}+\zeta_{q+j^*+2}-k-j^*} \ .
\EE
Recalling that $k>q$ and $j^*>q$ one has by Eq. (\ref{Holder})
\BE
\zeta_{k+j^*+2}-\zeta_{k+q+2} \leq \zeta_{q+j^*+2}-\zeta_{2q+2}\, .
\EE
Hence expression (\ref{eq7}) which coincides with the RHS of
Eq. (\ref{con}) is the leading one.
We have therefore proven that relation~(\ref{hyp}) implies
relation~(\ref{con}).
Now it remains to show that Eq. (\ref{hyp}) is true for $p=0$.
But this follows straightforwardly by setting $q=0$ in Eq. (\ref{eq1})
 and replacing $\k_{-1}$ by 1.
Hence assertion (\ref{sup}) is proven.
This in turn implies as was shown at the
beginning of this section that (\ref{kgs}) is satisfied.
\section{Scaling laws implied by the continued fraction representation:
  derivatives at time zero}
To prepare for the comparison between the
continued fraction and the multi-fractal representations we identify
in this Section the leading scaling exponents that characterize the
$n$th order time derivative of the correlation function
Eq. (\ref{expL}) at $\tau=0$.
We show that
\BE \label{result1}
\B.{\cal F}_2^{(n)}(R,0) \approx
 R^{\zeta_{2+n}-n} \;\;\;\;\forall n\ .
\EE
Here we use the shorthand notation
\BE
\B.{\cal F}_2^{(n)}(R,0) \equiv
{\partial^n\B.{\cal F}_2(R,\tau)\over \partial \tau^n}\Bigg|_{\tau=0} \ .
\EE
In the next Section we will
show that the same exponents are predicted by the
multi-fractal representation, making the prediction of the Taylor
expansion of the correlation function the same from the point of view
of scaling behavior.

From Eq.~(\ref{step0}) one deduces by inverse
Laplace transform the following  equation for
 $\B.{\cal F}_2^{(1)}(R,\tau)$
\BEA
\label{dot}
\B.{\cal F}_2^{(1)}(R,\tau)&=&\gamma_0(R) \B.{\cal
F}_2(R,\tau)\\ \nn
&&+\int_0^\tau \B.K_0(R,\tau')
\B.{\cal F}_2(R,\tau-\tau')d\tau' \ ,
\EEA
where $\B. K_0(R,\tau)$ is the inverse Laplace transform of
$\tilde \B. K_0(R,z)$ [Eq. (\ref{K0z})],
\BE
\label{K0rt}
\B. K_0(R,\tau)=\frac{1}{\B. k_0(R)}(\w_1,e^{\l_1 \tau}\w_1)\ .
\EE
Equation~(\ref{dot}) is the so-called memory-function equation, $\B.
K_0(R,\tau)$ being the memory kernel.
At $\tau=0$ Eq. (\ref{dot})
becomes
\begin{equation}
\B.{\cal F}_2^{(1)}(R,0)=\B.\gamma_0(R) \B. k_0(R)
\propto R^{\zeta_3-1} \ ,
\end{equation}
where we used the scaling laws (\ref{scalekg}).

The higher order partial time derivatives are  obtained by differentiating
 Eq.~(\ref{dot}),
\end{multicols}
\vskip -0.4cm
\leftline{----------------------------------------------------------------------
 --}
\begin{eqnarray}
\label{ndot}
\B.{\cal F}_2^{(n)}(R,\tau)=\B.\gamma_0(R)
\B.{\cal F}_2^{(n-1)}(R,\tau)
+\sum_{k=0}^{n-2} \B.K_0^{(n-k-2)}(R,\tau)
\B.{\cal F}_2^{(k)}(R,0)
+ \int_0^{\tau}  d\tau'    \B.K_0(R,\tau')
\B.{\cal F}_2^{(n-1)}(R,\tau-\tau')\ .
\end{eqnarray}
\vskip .1cm
\rightline{---------------------------------------------------------------------
 ---}
\begin{multicols}{2}
At $\tau=0$ one obtains
\begin{eqnarray}
\label{ndot0}
\B.{\cal F}_2^{(n)}(R,0)&=&\B.\gamma_0(R)
\B.{\cal F}_2^{(n-1)}(R,0) \\
&&+\sum_{k=0}^{n-2}\B.K_0^{(n-k-2)}(R,0)
\B.{\cal F}_2^{(k)}(R,0)\  .\nonumber
\end{eqnarray}
Now one realizes that this leads us to consider a hierarchy of equations for
$\B.K_{i}^{(q)}(R,0)$, $q \geq 1$, $i \geq 0$.
Noticing that Eq. (\ref{stepn+1}) is formally the same as
Eq. (\ref{step0}) one obtains in analogy with Eq. (\ref{dot})
\BEA
\label{redot}
\B.K_i^{(1)}(R,\tau)&=&\gamma_{i+1}(R)
\B.K_i(R,\tau)\\
&&+\int_0^\tau \B.K_{i+1}(R,\tau')
\B.K_i(R,\tau-\tau')d\tau' \ ,\nonumber
\EEA
where $\B.K_i(R,\tau)$ is the inverse Laplace transform of Eq. (\ref{Knz})
\BE
\label{Kirt}
\B.K_i(R,\tau)=(\tilde\w_{i+1},e^{\l_{i+1} \tau}\w_{i+1})(\tilde\w_i,\w_i)^{-1}\
 .
\EE
Notice that at $\tau=0$ using also Eq. (\ref{defk}) yields
\BE
\label{Kir0}
\B.K_i(R,0)=\B.k_{i+1}(R)\ .
\EE
In analogy with Eq. (\ref{ndot0}) one has for $q \geq 1$
\begin{eqnarray}
\label{rendot0}
\B.K_i^{(q)}(R,0)&=&\B.\gamma_{i+1}(R)
\B.K_i^{(q-1)}(R,0)\\
&&+\sum_{k=0}^{q-2\geq 0} \B.K_{i+1}^{(q-k-2)}(R,0)
\B.K_i^{(k)}(R,0)\  .\nonumber
\end{eqnarray}
Note that denoting from now on  $\B.K_{-1} \equiv \B.{\cal F}_2$ and
allowing $i \geq -1$ in Eq. (\ref{rendot0}) one recovers
Eq. (\ref{ndot0}).

It follows that to determine the scaling
behavior of $\B.K_{-1}^{(n)}(R,0)$ for $n \geq 1$ in order to prove
Eq. (\ref{result1}) one has to know $\B.K_{i}^{(q)}(R,0)$ $\forall$
$q+2i \leq n-2$
 with $q\geq 0$ and
$i \geq -1$.
It is sufficent to prove the following
\BE
\label{hyp2}
\B.K_{i}^{(q)}(R,0) \propto R^{\zeta_{q+2i+4}-\zeta_{2i+2}-q-2+2
\delta_{i,-1}} \  ,
\EE
where $q \geq 0$, $i \geq -1$ and $\delta_{i,-1}$ is a Kronecker
delta, which we do by induction
in the sequel.
Setting $i=-1$ and recalling that $\zeta_0=0$ Eq. (\ref{result1}) results
 directly.

Let us assume that Eq. (\ref{hyp2}) is true
$\forall \  q+2i \leq n-3$ with  $n \geq 1$.
Then we show that this expression is still valid for $n+1$.
By the scaling relations (\ref{scalekg}) we already know that
Eq. (\ref{hyp2}) is true
$\forall \  q+2i \leq n-3$ with  $n=1$.
For $q+2i=n-2$ Eq. (\ref{rendot0}) becomes
\begin{eqnarray}
\label{rerendot0}
\B.K_i^{(n-2-2i)}(R,0)&&=\B.\gamma_{i+1}(R)
\B.K_i^{(n-3-2i)}(R,0)\\
+\sum_{k=0}^{n-4-2i \geq 0}&&\B.K_{i+1}^{(n-4-2i-k)}(R,0)
\B.K_i^{(k)}(R,0)\  .\nonumber
\end{eqnarray}
Note that for $2i=n-2$ ($n$ even) one has to resort to
Eq. (\ref{Kir0}) and that for $2i=n-3$ ($n$ odd) the sum in
Eq. (\ref{rerendot0}) is empty yielding  directly
\BEA
\label{kspe}
&&\B.K_{\frac{n-2}{2}}(R,0)=\B.k_{\frac{n}{2}}(R)\propto
R^{\zeta_{n+2}-\zeta_{n}-2}\ ,\\
&&\B.K_{\frac{n-3}{2}}^{(1)}(R,0)=
\B.\gamma_{\frac{n-1}{2}}(R)\B.k_{\frac{n-1}{2}}(R)\propto
R^{\zeta_{n+2}-\zeta_{n-1}-3+2
\delta_{n,1}}\ ,\nonumber
\EEA
respectively, which are indeed of the form of Eq. (\ref{hyp2}).
Now for $n-4-2i \geq 0$ the only object in Eq. (\ref{rerendot0}) whose
scaling is not known by Eq. (\ref{hyp2}) for $q+2i\leq n-3$ is that
corresonding to $k=0$, $\B.K_{i+1}^{(n-4-2i)}(R,0)$.
Hence we prove also by induction, but now on $i$,
that Eq.(\ref{hyp2}) holds for $q+2i=n-2$.
Assuming that $\B.K_{i+1}^{(n-4-2i)}(R,0)$  scales according to Eq. (\ref{hyp2})
 we
show that it is also true for $i-1$.
By Eq. (\ref{kspe}) we already  know that this is true for  $2i=n-4$ ($n$
even) and $2i=n-3$ ($n$ odd).
We first show that for $n-4-2i \geq 0$ the term in $k=0$ in
Eq. (\ref{rerendot0}) is leading with respect to the terms in higher
$k$ appearing in this equation.
By hypothesis eq. (\ref{rerendot0}) yields
\BEA
\label{eqs}
&&\B.K_i^{(n-2-2i)}(R,0) \propto
R^{-\zeta_{2i+4}-\zeta_{2i+2}-n+2i+2
\delta_{i,-1}}\\
&&\left(
R^{\zeta_{2i+5}+\zeta_{n+1}}+\sum_{k=0}^{n-4-2i \geq 0}
R^{\zeta_{n+2-k}+\zeta_{2i+4+k}}\right) \ .\nonumber
\EEA
Hence one has to show that for $k > 0$
\BE
\label{j'ai}
\zeta_{n+2}-\zeta_{n+2-k} \leq \zeta_{2i+4+k}-\zeta_{2i+4} \ ,
\EE
which follows from the H\"older inequalities since $k \leq n-4-2i$.
Secondly, we notice from Eq. (\ref{eqs}) that the term in front of the
sum  scales exactly as a term in $k=1$ so that by the preceding
argument it is subleading with respect to the term in $k=0$ for $n-4-2i
\geq 0$.
It follows that the term in $k=0$ is the leading one in
Eq. (\ref{eqs}) so that coming back to Eq. (\ref{rerendot0}) one has
\BEA
\label{faim}
\B.K_i^{(n-2-2i)}(R,0) &\approx&
\B.K_{i+1}^{(n-4-2i)}(R,0)\B.K_i^{(0)}(R,0) \\
&\propto& R^{\zeta_{n+2}-\zeta_{2i+2}-n+2i+2
\delta_{i,-1}}\ . \nonumber
\EEA
Hence  Eq. (\ref{hyp2}) is true for $q+2i=n-2$
with $q \geq 0$ and  by induction on $i$ for any  $i \geq -1$.
Now by induction on $n$ it  is also  true for any  $n$.
We have therefore proven Eq. (\ref{hyp2}) and henceforth
Eq. (\ref{result1}).
As for Eq. (\ref{result1}) it results directly from Eqs. (\ref{kspe}) and
 (\ref{faim})
\section{The multi-scaling representation}
The multi-scaling representation of $\B.{\cal F}_2(R,\tau)$ can be written
as \cite{97LPP}:
\BE
\B.{\cal F}_2(R,\tau)=U^2\int d\mu(h) \Big(\frac{R}{L}\Big)^{2h+{\cal Z}(h)}
\B.f_2\Big(\frac{\tau}{\tau_{R,h}}\Big)\,,
\label{Fth}
\EE
where $U$ is the characteristic magnitude of the velocity difference
across the outer scale of turbulence, $\B.f_2$ is a function of
the scaled time variable only, and
\BE
\tau_{R,h}\sim \frac{R}{U}\Big(\frac{L}{R}\Big)^h \ .
\EE
The function ${\cal Z}(h)$ is related to the scaling exponents $\zeta_n$
of the $n$th order structure functions through the saddle point requirement
\BE
\zeta_n=\min_h [nh+{\cal Z}(h)]\ . \label{minh}
\EE
To find the scaling exponents associated with the time derivatives of
$\B.{\cal F}_2(R,\tau)$ at $\tau=0$ one computes the $n$-th order
partial time derivative of Eq.~(\ref{Fth}) to obtain
\BEA\label{ndotFth}
&&\B.{\cal F}_2^{(n)} (R,\tau)\\ \nn
& =& \frac{U^{2+n}}{R^{n}}\int
 d\mu(h) \Big(\frac{R}{L}\Big)^{(2+n)h+{\cal Z}(h)}
\B.f_2^{(n)}\Big(\frac{\tau}{\tau_{R,h}}\Big)\,,
\EEA
where
\BE
\B.f_2^{(n)}(s)={d^n \B.f_2(s) \over d s^n}\ .
\EE
At $\tau=0$ this gives
\BE
\B.{\cal F}_2^{(n)}(R,0)=\B.f_2^{(n)}(0)\frac{U^{2+n}}{R^{n}}\int d\mu(h)
 \Big(\frac{R}{L}\Big)^{(2+n)h+{\cal Z}(h)}\ .
\label{ndotFth0}
\EE
Computing the integral at the saddle point in the limit $R/L\to 0$
and using (\ref{minh}) we find
\BE
\B.{\cal F}_2^{(n)}(R,0)\propto
R^{\zeta_{2+n}-n}\,,
\EE
in correspondence with Eq.~(\ref{result1}). We thus see that the
continued fraction representation generates the same predictions regarding
the multiplicity of time scales characterizing the time correlation
functions as the multi-scaling representation. We take this as
an independent evidence for the correctness of the latter.
\section{conclusion}
In conclusion, we showed that the formally exact continued
fraction representation of the time correlation functions of
Belinchier-L'vov velocity differences has the same Taylor expansion as the
multi-scaling representation, at least in terms of the leading scaling
exponents order by order.

It should be noted that the continued fraction representation can be
used as an approximant for the correlation function when analytic
forms of the time-correlation functions are needed.  In the lowest
approximation one takes in Eq.~(\ref{contfrac}) $\B.k_1(R)=0$,
producing an exponential decay of the correlation function, with a
typical decay rate $\gamma_0(R)$. In every successive approximation
($\B.k_2=0$, $\B.k_3=0$, etc.)  one introduces more and more
characteristic scales, each one characterized by a different
``dynamical exponent", taking progressively more
information about the statistics of higher order correlation functions
into account.

\acknowledgements
We thank Daniela Pierotti and Reuven Zeitak for discussions concerning
the non-Hermiticity
of the Liouvillian. This work was supported in part by the German Israeli
Foundation, the European Commission under the Training
and Mobility of Researchers program, The Basic
Research Fund administered by the Israel Academy of Science and
Humanities, the Minerva Center for Nonlinear Physics, and the Naftali
and Anna Backenroth-Bronicki Fund for Research in Chaos and
Complexity.

\end{multicols}
\end{document}